# nanoHUB services for FAIR simulations and data: ResultsDB and Sim2Ls

Daniel Mejia, Steven Clark, Juan Carlos Verduzco, Michael Zentner, Lynn Zentner, Gerhard Klimeck, Alejandro Strachan[*]

**Keywords:** simulation, cyber-infrastructure, multiscale modeling, data science

nanoHUB is an open cyber platform for online simulation, data, and education that seeks to make scientific software and associated data widely available and useful. This paper describes recent developments in our simulation infrastructure to address modern data needs. nanoHUB's Sim2Ls (pronounced sim tools) make simulation, modeling, and data workflows discoverable and accessible to all users for cloud computing using standard APIs. In addition, published tools are findable (with digital object identifiers), reusable (via documented requirements and services), and reproducible via containerization. In addition, all Sim2L runs are automatically cached, and their results indexed into a global and queryable database (ResultsDB). We believe this infrastructure significantly lowers the barriers towards making simulation/data workflows and their data findable, accessible, interoperable, and reusable (FAIR). This frictionless access to simulations and data enables researchers, instructors, and students to focus on the application of these products to advance their fields.

---

[*] Corresponding author, Email: strachan@purdue.edu.

## 1. Introduction

Data science and physics-based simulations are playing an increasingly important role in engineering and the physical sciences[1]. Maximizing the impact of modeling and simulation requires making the associated tools and data generated widely accessible and useful. In the case of research data, successful efforts toward making them findable, accessible, interoperable, and reusable (FAIR) [2] can enable data-science-driven research. Despite these resources and efforts to improve data sharing at national and international levels, much of the research data generated remains inaccessible. Progress is also occurring toward increasing access to simulations and models. Examples include platforms for online simulation and modeling [3], easy-to-use libraries for machine learning, and materials property databases. Despite this progress, scientific software often uses *ad hoc* input languages with arbitrary ontologies and typically unverified input values. Each tool has its own idiosyncrasy and requires a significant learning curve. This tends to limit the use of these tools to relatively small groups of experts and makes reproducibility a tall order. These limitations are compounded by the fact that most computational research involves complex multi-step workflows which are often carried out using poorly documented single-use scripts or undocumented manual steps. Even when workflows are shared electronically, reconstructing the simulation environment is often challenging and time-consuming. In most cases, publishing or sharing the specific workflows used in research and the data generated remains a tedious and time-consuming afterthought. Recent efforts designed to document and share scientific workflows [4], [5] and data management [6] are beginning to address these challenges and this paper describes infrastructure developments at nanoHUB designed to address this gap.

nanoHUB's mission is to make scientific software and associated data available and useful to the research, development, and education communities. The platform has removed many of the traditional barriers, including computational expertise, access to specialized hardware, and software-specific knowledge[7]. Over 165,000 simulation users have performed millions of online simulations. Lowering barriers also broadens participation; users from 140 minority-serving institutions in the US have performed simulations in nanoHUB. Scientific software developers can use nanoHUB to transform their products into containerized Apps and tools that run in the cloud with easy-to-use interfaces presented through a web browser, see Figure 1. Published nanoHUB tools are assigned digital object identifiers and are indexed by Web of Science, providing developers a formal recognition for sharing their work. This translational process turns computational research outcomes into products that can be used by experimentalists, instructors, and students. Thus, domain experts do not need to worry about computational aspects or software idiosyncrasy and can focus on the application of scientific software to advance their fields. This paper focuses on recent advances in nanoHUB simulation delivery ecosystem, Sim2Ls, and the results database (ResultsDB), that make scientific workflows and their data FAIR. Developers can share their workflow as a published Sim2L in nanoHUB, any user can execute published Sim2Ls, and all the community-generated data with the open resources is automatically indexed and made available.

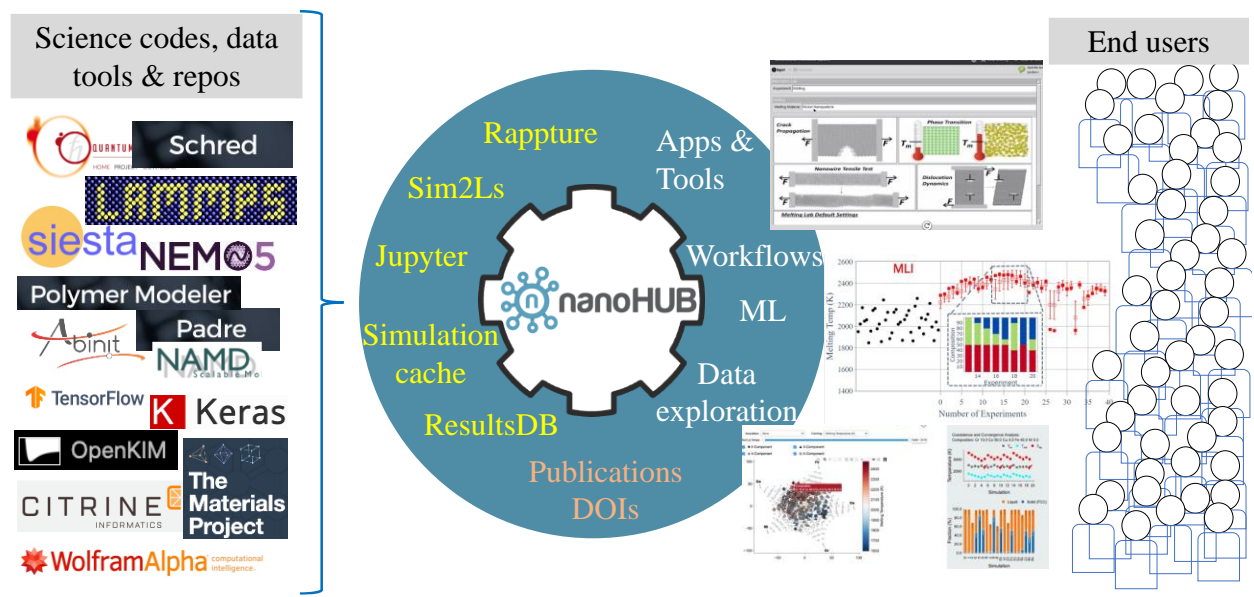

*Figure 1. nanoHUB simulation ecosystem connects science codes and data science tools to end-users via easy-to-use online Apps, workflows, and tools.*

## 2. Results

nanoHUB supports the deployment and publication of various types of tools and Apps. These include Apps created using Rappture[8], Jupyter notebooks, codes with X11 graphical interfaces, and commercial software. These have been described in prior publications; here we focus on the infrastructure for FAIR workflows and data.

**FAIR workflows and data.** With the growing importance of data-driven research, there is a clear need for FAIR scientific workflows and data. To address this challenge, we introduced Sim2Ls. With Sim2Ls, developers use modern tools to define end-to-end, reproducible simulation or analysis tasks with verified inputs and outputs[5]. Sim2Ls are Jupyter notebooks and must include a declaration of inputs and outputs using YAML as well as the required computational workflow. The latter should include all pre- and post-processing steps and can involve computationally intensive simulations submitted to HPC resources. Inputs have well-defined types, descriptions (metadata), and, optionally, units and ranges. At run time, the Sim2L library verifies that all inputs satisfy requirements, performs automatic unit conversion, and executes the workflow. After execution, output variables are checked for completeness and the executed notebook and auxiliary files are returned to the requesting user. As described below in more detail, results from successful runs are automatically indexed and stored in a globally accessible database, the ResultsDB.

Published Sim2Ls are accessible via Apps or workflows (Jupyter or command line) from within nanoHUB or invoked from outside the platform via Web Services. Sim2Ls, their services (outputs), and requirements (inputs) are also findable. Sim2L names, descriptions, inputs, and outputs can be queried using an API [9]. Providing detailed metadata for all inputs and outputs is critical to making the tool, its services, and its requirements FAIR.

A recent publication [10] exemplifies the use of a Sim2L that calculates the melting temperature of high-entropy alloys using molecular dynamics, invoked from an autonomous active learning workflow to find the alloy composition with the highest melting temperature. Both the underlying Sim2L and the active learning workflows are published and accessible. As will be described below, the ResultsDB enables the exploration of all the underlying data used in the publication and that generated since then. We note that Sim2Ls are not limited to computational work, they can also be used to analyze experimental data and make data and analysis FAIR.

**Publishing tools.** Through nanoHUB's tool publication mechanism, developers can make their work broadly available. These products can be developed independently or using nanoHUB's software development environment, see Methods. The publication process begins with the registration of a tool and the selection of a repository to manage the software. The tool is then uploaded and tested in nanoHUB's environment. Once testing is complete and the tool is approved for publication by the developer, it is published and assigned a DOI. Developers can continue to update and improve their tools and publish new versions, each with their own DOI to preserve the provenance of the code.

**Simulation cache and results retrieval.** Every Sim2L or Rappture simulation performed in nanoHUB is automatically stored in a simulation cache for future retrieval. Using hashing to enable fast identification of identical run requests, nanoHUB provides results instantaneously if a user requests a previously performed simulation. Instantaneous feedback not only enables more efficient use of computationally intensive simulations in education and speeds up research but also impacts resource use.

nanoHUB keeps track of the jobs performed for users and the runs delivered by the cache. Many Rappture and Sim2L tools involve workflows that require multiple individual jobs, which are individually counted, and we also keep track of runs that represent the entire workflow. Figure 2 (a) shows the number of simulation jobs performed for users as a function of time, and we show cumulative jobs for the 12 months trailing. Since the creation of nanoHUB, these numbers have increased steadily until we started testing the caching system. In 2019 the production version of the cache was deployed and a large increase in runs delivered by the cache followed, and Figure 2 (b) shows 12-month trailing cache runs. The deployment of the cache resulted in a temporary decrease in the jobs performed, Figure 2(a), freeing up computational resources for other tasks. The increase in popularity of Jupyter for interactive computing, including machine learning, in 2020 resulted in an increase in the number of jobs performed for users (which include Jupyter sessions).

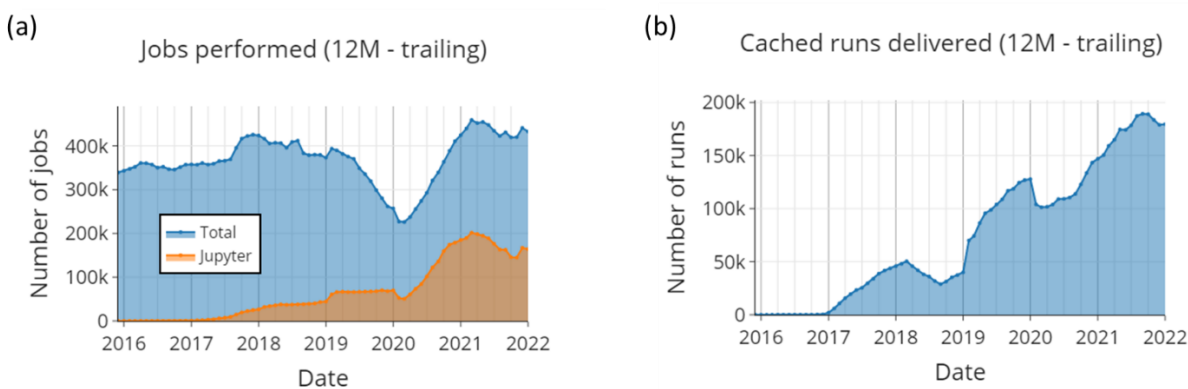

*Figure 2. Simulation stats. Simulation jobs performed for the users (a) and runs delivered by the cache (b).*

**ResultsDB.** Making research data FAIR is often a burdensome afterthought, even in the case of computational work. To turn this into a seamless process, the nanoHUB simulation ecosystem indexes the outputs from Rappture and Sim2L runs into a globally accessible results database (ResultsDB). As of January 2023, the ResultsDB contains 1.26+ million entries from Rappture tools and 12,700+ for the newly deployed Sim2Ls. Users can explore the database using a web interface or query its results using an API[9] to learn from community results before performing their own runs.

The web interface is available at [https://nanoHUB.org/results](https://nanoHUB.org/results), see Figure 3, and enables users to find both Rappture and Sim2L datasets using various filters. Each dataset (associated with its underlying tool) has a DOI, Fig. 3 (a), and metadata, see Fig. 3(b), including a schema associated with inputs and outputs in a Sim2L. The web interface shows the schema of the latest version of the tool and the API provides access to the schema of prior versions. Each data entry within each dataset has a unique internal identifier, see Fig. 3(c). The web component consists of three controllers: *i) browse*, to explore datasets and their metadata; *ii) results*, to show data and its metadata; *iii) download*, to access any of the research objects included in the dataset. Data is displayed as a table where all inputs and outputs are visible, each record includes its serialized query identifier (SQuID)and the creation date. Inputs and outputs of a basic type (e.g., text and numbers) are directly displayed on the table, and links are included for more complex types (e.g., lists, arrays, or images). Whenever possible, pages include code snippets to obtain the same results using python via the nanoHUB-remote library, but the same results can be obtained using HTTP.

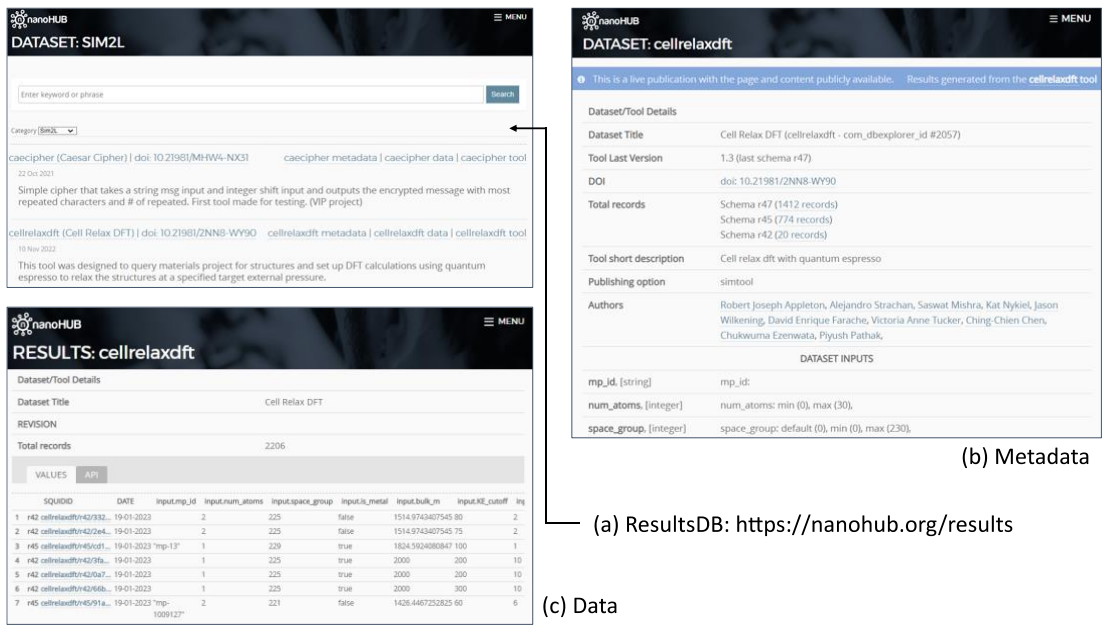

*Figure 3. Web interface to the ResultsDB. (a) Access point listing Sim2L and Rappture entries in the database; each dataset has an associated DOI. (b) The metadata for each tool includes versions, the input-output schema. (c) Data entries associate with a Sim2L every entry has a unique internal identifier. In addition to the Web interface all the information associated with the ResultsDB is available via an API.*

In addition to the web interface, the ResultsDB can be accessed via a REST API. This allows users to request detailed information about schemas and datasets generated by tools and subset of the results generated by Sim2Ls and Rappture tools. The API receives a request with specifications and filters and returns a list of matching inputs and outputs. The API also allows users to request basic statistics for results matching a collection of restrictions, e.g., median or standard deviation. The API uses OAuth authentication protocol via nanoHUB credentials or session tokens (https://nanohub.org/api/results/dbexplorer/) .

**Data exploration and simulation tools.** The combination of simulation as a service and the ResultsDB enables new ways of interacting with simulations and their results. In this new paradigm, an app launches with preloaded results from the ResultsDB instead of a set of input parameters. Users can visually explore and learn from community results, find unexplored spaces, and perform additional simulations as needed, see Figure 4.

An example of such Apps is the PNJunction Lab Exploration tool (doi:10.21981/3CJ3-6F56), top row of Figure 4. This app loads current-voltage data from the PNjunction tool (doi:10.4231/D3GH9B95N) stored in the ResultsDB. Users define device dimensions and the material of interest using the graphical elements of the exploration tool and the app displays available results in terms of the acceptor and donor doping, as shown in the top row of Figure 4. This matrix classifies results by orders of magnitude in doping, this visualization helps users to quickly understand trends on device performance. If there are no results on the database, the corresponding spot on the matrix is empty, and users can launch new simulations to explore the uncharted parameter space. The tool uses all the parameters defined by the user to launch a new simulation to the PNjunction tool using nanoHUB's API. A second example explores results obtained with a DFT tool (doi:10.4231/D30G3H12Q) that computes electronic and mechanical properties of a wide range of materials. The DFT Results Explorer tool (doi:10.21981/ATQY-0Z53) presents these results with a user-friendly visualization enabling the exploration of how various inputs (e.g. kinetic energy cutoff or number of k-points) affect predictions. Again, users can request additional simulations to be performed directly from the data exploration tool. These are two examples of what can be done, but any nanoHUB user can develop their own data exploration tool connected to any of the Rappture or Sim2Ls currently available.

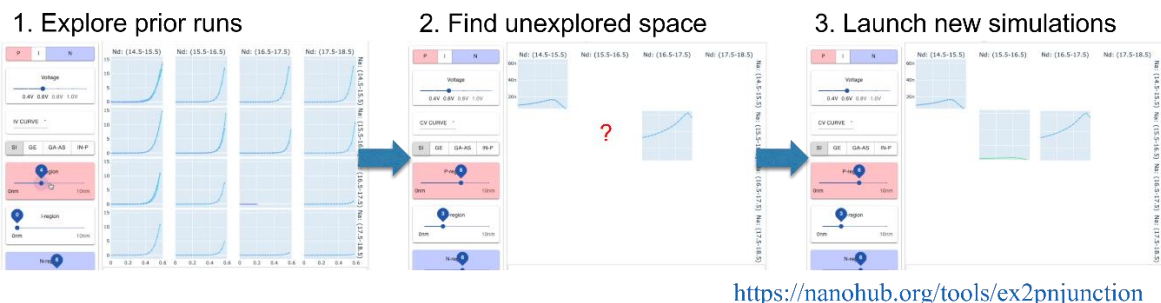

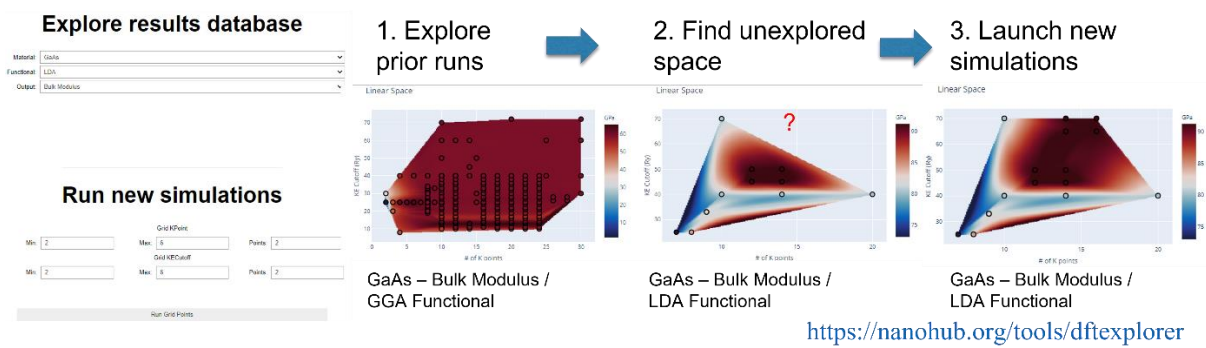

*Figure 4. The PNJunction Lab Exploration Tool (top) allows users to visually explore device's properties for multiple doping concentrations, users can modify the input parameters to explore different configurations of the device. The DFT Results Explorer Tool (bottom) presents a user-friendly representation of calculated electronic and mechanical properties of a wide range of materials.*

Figure 5 demonstrates another use of the ResultsDB, in this case to create an interactive dashboard to explore existing computational simulation results. The MeltDashboard tool (doi:10.21981/A6GT-V318) queries results of molecular dynamics simulations of the melting temperature of high entropy alloys, using the MeltHEAs Sim2L (doi:10.21981/ER9D-MM30). The tool maps the alloys composition into two dimensions and displays the melting temperature. Users can click on a specific alloy and explore individual simulations (right panel). In addition, users can perform simulations on regions of composition space that have not been explored fully.

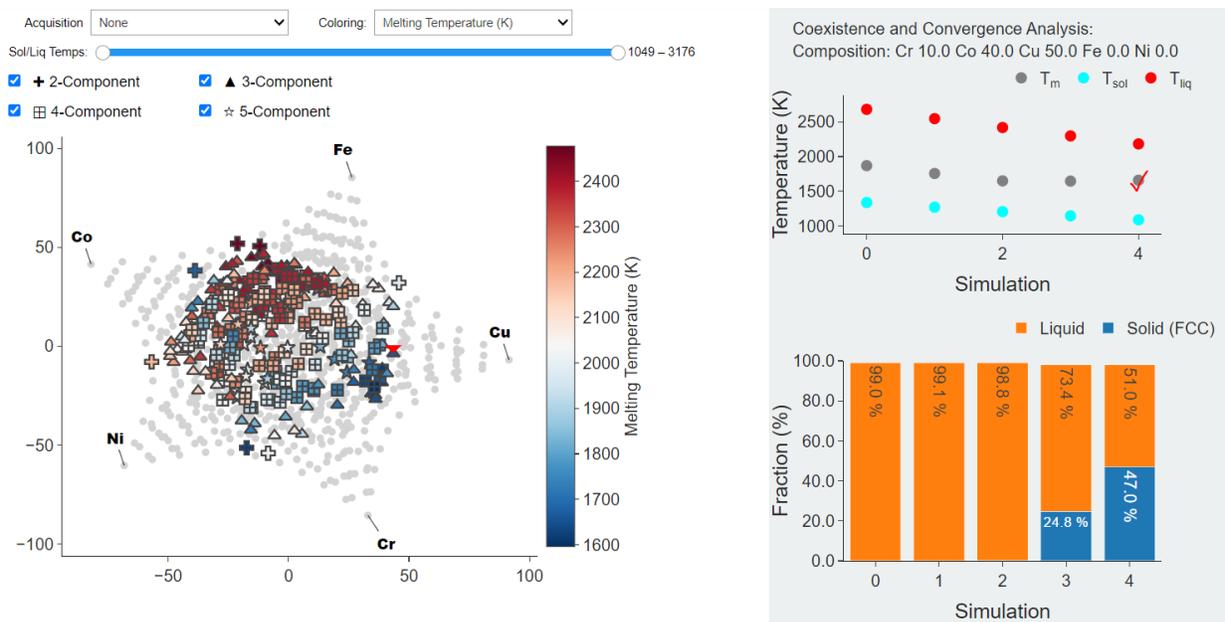

*Figure 5. The MeltDashboard Tool allows users to visualize alloy melting temperature results from MD simulations stored in the ResultsDB. Users can interact with the interface to explore different compositions and explore the simulations required for a converged melting temperature.*

## 3. Discussion

Realizing the full potential of data science and physics-based simulations in the physical sciences and engineering requires making research data and simulation/analysis workflows FAIR. nanoHUB is an end-to-end platform to develop, publish, and consume simulations, data, and associated educational/training material. Software developers can build on a powerful software environment and their tools can benefit from features like automatic uncertainty quantification[11], libraries to streamline the development of user interfaces, simulation result caching, and automatic output data indexing. Recent developments, Sim2Ls and the ResultsDB, enable computational researchers to share their products more easily, improve reproducibility, and provide an avenue to make computational workflows and their data FAIR. These tools can also be used by experimentalists to analyze and share their data. We believe that these new ways of sharing and consuming simulations and data will accelerate innovation in research, education, and workforce development.

## 4. Methods

Figure 6 provides an overview of nanoHUB's infrastructure, based on the HUBzero® framework, including selected external partners. End-users consume simulation and data using Apps, workflows, and data exploration tools. The content management system (CMS) exposes the various services to users who can interact with simulations via Jupyter notebooks, in-house Rappture[8] and Sim2Ls libraries[5], or directly with X11 applications. Under the hood, single simulations, parallel jobs, and UQ workflows are orchestrated by a series of engines that provide postprocessing services, simulation caching, and submission to various computing resources. A variety of computational resources can be accessed, including campus clusters, XSEDE resources, and high throughput computing (HTC) resources like Open Science Grid (OSG) and BOINC [12]. In addition, nanoHUB supports highly interactive sessions for computation, visualization, and machine learning with local "instantly available" resources with both CPUs and GPUs.

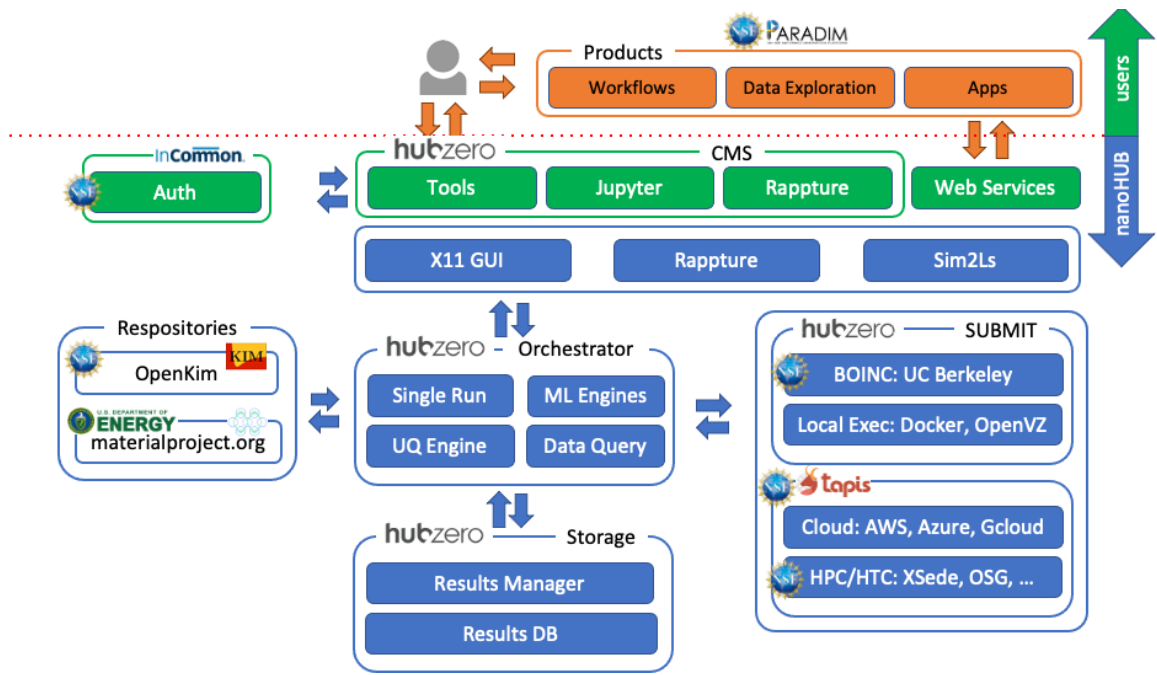

*Figure 6. Overview of nanoHUB's platform, including a user-facing CMS and back-end tools and workflows for computing.*

nanoHUB provides three complementary environments for developers to create and maintain applications: i) Linux workspace, ii) Jupyter notebook environment, and iii) RStudio environment. The user interfaces for these three environments are drastically different and support different development paradigms while sharing common features rooted in the underlying Linux operating system.

## 5. Data Availability

Datasets described in this paper are available in nanoHUB.org through open-source tools under GNU General Public licenses. The ResultsDB, which hosts data obtained through runs in nanoHUB, is hosted in https://nanohub.org/developer/api/endpoint/dbexplorer under a Creative Commons CC BY 4.0 license. It is accessible through the nanoHUB remote [9] library inside nanoHUB.

## 6. Code Availability

The source code for nanoHUB and nanoHUB services is built over the HUBzero® framework, which was released under an MIT license and is made available in https://hubzero.org/services/opensource. Source code for Rappture is available in rappture.org under an MIT license. Source code for the Sim2lBuilder library is available in https://pypi.org/project/sim2lbuilder/. Code necessary for Papermill execution is hosted on https://pypi.org/project/papermill/ under a BSD license. Source code for the tools featured in this paper is available on nanoHUB.org.

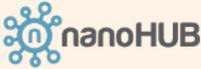

## Acknowledgments

This work was partially supported by the Network for Computational Nanotechnology, a project of the US National Science Foundation, EEC-1227110. J. C. V. thanks the Science and Technology Council of Mexico (CONACYT) for partial financial support of this work. The authors would like to acknowledge Mr. Travis Merrick for data analysis.

## Author Information

### Affiliations

**Network for Computational Nanotechnology, West Lafayette, IN, USA**

Alejandro Strachan, Juan Carlos Verduzco, Gerhard Klimeck, Daniel Mejia, Lynn Zentner

**School of Materials Engineering, Purdue University, West Lafayette, IN, USA**

Alejandro Strachan, Juan Carlos Verduzco

**San Diego Supercomputer Center, University of California, San Diego, USA**

Steven Clark, Michael Zentner

**School of Electrical and Computer Engineering, Purdue University, West Lafayette, IN, USA**

Gerhard Klimeck

### Contributions

A.S., G.K., and Mark S. Lundstrom conceived the simulation and data services; S.C., D.M., A.S. architected and S.C. and D.M. implemented the services. J.C.V. and D.M. developed the user-facing products. G.K. and L.Z. lead the analytics work. M.Z. leads the operation of the platform. A.S., S.C., D.M., and J.C.V. wrote the paper. All authors edited the paper. Early platform developments are described in [8].

## Corresponding Authors

Correspondence to Alejandro Strachan (strachan@purdue.edu)

## Ethics Declarations

### Competing Interests

The authors declare no competing interests.

## Author Biographies

***Dr. Alejandro Strachan*** is a Professor of Materials Engineering at Purdue University. He received a Ph.D. in Physics from the University of Buenos Aires, Argentina. Prof. Strachan's research focuses on the development of predictive atomistic and multiscale models to describe materials from first principles and their combination with data science to address problems of technological or scientific importance. He is known for his work on atomic-level modeling of materials at extreme conditions of pressure, temperature, and deformation rates. He has published over 180 papers in these areas. Contact him at strachan@purdue.edu

**Dr. Steven Clark** is currently a Computational and Data Science Research Specialist with the San Diego Supercomputer Center. He received a PhD in Chemical Engineering from Purdue University. Dr. Clark joined the nanoHUB team more than 15 years ago doing work in tool deployment and pipeline development with particular focus on use of HPC and HTC resources. Contact him at smclark@sdsc.edu.

**Mr. Juan Carlos Verduzco** is a PhD candidate in the School of Materials Science and Engineering at Purdue University. He holds a B.S. in Mechatronics from the Monterrey Institute of Technology and Higher Education, Mexico. His research is focused on coupling experimental techniques, physics-based simulations, and machine learning algorithms for the rational design of novel battery materials. Contact him at jverduzc@purdue.edu

**Dr. Daniel Mejia** is a Senior Software Developer at the Network for Computational Nanotechnology at Purdue University. He has more than 20 years of experience delivering solutions to industry and education. He served as Chief Technology Officer, Dayscript/grupo-link Colombia and was lecturer at Universidad de los Andes, Bogota, Colombia. Dr Mejia received his Ph.D. in Electrical and Computer Engineering from Purdue University. Contact him at denphi@denphi.com.

**Dr. Lynn K. Zentner** is the Managing Director of the Network for Computational Nanotechnology. She has served as Chief Operating Officer and Senior Vice President, CBIOS Corporation, and was previously a visiting professor in the Mechanical Engineering Technology Department at Purdue. Dr. Zentner received a PhD in Mechanical Engineering from Purdue University and has been involved in modeling and simulation research for over 20 years. Contact her at lzentner@purdue.edu

**Dr. Michael G. Zentner** is the Division Director for Sustainable Scientific Software at the San Diego Supercomputer Center at University of California, San Diego. He leads the Science Gateways Community Institute to grow the national community around science gateways, the Hubzero project to provide an easy-to-use software framework for building science gateways, and the Rev-Up program to assist researchers in creating sustainability models for their projects. Previously, he has been a founder/senior team member of several technology companies, where he created innovative solutions for extracting patterns from data, collaboration tools, and supply chain optimization systems**.** Contact him at mzentner@ucsd.edu

**Dr. Gerhard Klimeck** is an Electrical and Computer Engineering faculty at Purdue University and leads two research centers in Purdue's Discovery Park. He helped to create nanoHUB.org which now serves over 2.0 million users globally. Previously he worked with Texas Instruments and NASA/JPL/Caltech. His research interest is in computational nanoelectronics, high performance computing, and data analytics. NEMO, the nanoelectronic modeling software built in his research group established the state-of-the-art in atomistic quantum transport modeling. He published over 525 printed scientific articles that resulted in over 20,000 citations and an h-index of 69 in Google scholar. Contact him at gekco@purdue.edu